\documentclass[prl,twocolumn,showpacs,amsmath,superscriptaddress,amssymb]{revtex4}

\usepackage{graphicx}
\usepackage{dcolumn}
\usepackage{bm}

\begin{document}


\title{Interfacial Fermi Loops from Interfacial Symmetries}

\author{Ryuji Takahashi}
 \affiliation{Department of Physics, Tokyo Institute of Technology,
2-12-1 Ookayama, Meguro-ku, Tokyo 152-8551, Japan}
 \affiliation{Department of Applied Physics, University of Tokyo,
7-3-1 Hongo, Bunkyo-ku, Tokyo 113-8656, Japan}

\author{Shuichi Murakami}
\affiliation{Department of Physics, Tokyo Institute of Technology, 
2-12-1 Ookayama, Meguro-ku, Tokyo 152-8551, Japan} 
\affiliation{TIES, Tokyo Institute of Technology, 2-12-1 Ookayama, Meguro-ku, Tokyo 152-8551, Japan}

\date{\today}

\begin{abstract}
%
We propose a concept of interfacial symmetries such as interfacial particle-hole 
symmetry and interfacial time-reversal symmetry, which appear in interfaces between two regions related to each other by particle-hole or time-reversal transformations. 
These symmetries result in novel dispersion
of interface states. In particular for the interfacial particle-hole symmetry the gap closes
along a loop (``Fermi loop") at the interface. We numerically demonstrate this
for the Fu-Kane-Mele tight-binding model. We show that the Fermi loop 
originates 
from a sign change of a Pfaffian of a product between the Hamiltonian and 
a constant matrix. 
\end{abstract}

\pacs{73.20.-r,75.70.Tj, 71.70.Ej}
\maketitle 
%

Recently, various 
physical phenomena originating from topological metallic states at the boundary \cite{Kane05a,Moore07,Fu07,Hasan10} have been studied.
In a topological insulator (TI) the bulk is gapped, while there are gapless states in its boundary.
The gapless surface states are topologically protected by time-reversal symmetry (TRS),
and typically show massless Dirac cones, as has been observed
experimentally in Bi$_2$Se$_3$ and Bi$_2$Te$_3$
 \cite{Xia09,Hsieh09a,Hsieh09b,Chen09}.
They are characterized by $Z_2$ topological numbers. 
In addition, there are also 
other topological phases protected by spatial crystallographic symmetries. 
In particular, the mirror Chern number \cite{Teo08}
characterizes topological phases in 
mirror-symmetric systems, and leads to gapless surface states 
\cite{Teo08,THsieh12} and interface states 
\cite{Takahashi11}.
We predicted that gapless interface states appear when
two TIs with opposite signs of the Dirac-cone chiralities are attached together \cite{Takahashi11}.
These gapless states originate from the difference of the mirror Chern numbers between the two TIs \cite{Takahashi11}.
%
The resulting interface states are trivial in terms of the $Z_2$ topological number, but protected by mirror symmetry.
%

In this paper, 
we propose interfacial symmetries and show resulting peculiar gapless interface states by 
these symmetries. For illustration, it is demonstrated 
for the Fu-Kane-Mele (FKM) tight-binding model \cite{Fu07}, 
with 
 an interface between two regions with opposite mirror Chern numbers, 
i.e. opposite signs of the spin-orbit coupling (SOC). 
In the absence of interfacial symmetry, the gapless interface states
due to the mirror Chern number 
consist of Dirac cones.
On the other hand, when the SOC parameters in the two regions have the same magnitude
with opposite signs, 
the whole system has particle-hole symmetry (PHS) at the interface, and 
anomalous gapless interface states appear. In these gapless states, the gap closes along a loop (``Fermi loop") in two-dimensional (2D) $k$ space, which is totally unexpected for
lattice models without any continuous rotational symmetry.
This Fermi loop appears in general interface systems, consisting of two regions related to each other by the PHS, even when the two regions are gapped.
We call this interfacial symmetry {\it interfacial particle-hole symmetry (IPHS)}.
In the presence of this new symmetry, the degeneracy is shown to occur along a loop. 
Similar discussion for {\it interfacial time-reversal symmetry (ITRS)} is also given.
Finally, we show the Fermi loops in the $\pi$-junction interface between the two Rashba systems as another example of the IPHS.


The FKM tight-binding Hamiltonian on the diamond lattice is represented as
\begin{eqnarray}
H&=&\sum_{\langle ij\rangle}t_{ij}c^{\dagger}_{i}c_{j}
+
\frac{8\lambda i}{a}\sum_{\langle \langle ij\rangle\rangle}
c^{\dagger}_{i}\mathbf{s}\cdot(\mathbf{d}_{ij}^{1}\times\mathbf{d}_{ij}^{2})c_{j}.\label{eq:FKMH}
\end{eqnarray}
$c_{i}$ ($c^{\dagger}_{i}$) is the annihilation (creation) operator of an electron at $i$-th site.
$t_{ij}$, $\lambda$ are real constants and $a$ is the lattice spacing.
The first term describes the nearest-neighbor hopping, which is independent of spin. The second term is the next-nearest-neighbor hopping term, which represents the SOC. $\mathbf{s}$ is the Pauli matrices for spins, and $\mathbf{d}_{ij}^{1,2}$ are two vectors constituting the next-nearest bond from the site $i$ to $j$.
We define the nearest-neighbor bond vectors $\boldsymbol{\tau}_{i=1-4}$ as
$\boldsymbol{\tau}_{1}=\frac{1}{4}(1,1,1)$,$\boldsymbol{\tau}_{2}=\frac{1}{4}(-1,1,-1)$, 
$\boldsymbol{\tau}_{3}=\frac{1}{4}(-1,-1,1)$, and $\boldsymbol{\tau}_{4}=\frac{1}{4}(1,-1,-1)$.

In the FKM model we consider a (111) interface between two regions of the 
FKM model, which have different signs of the SOC.
We impose a mirror symmetry with respect to the $(1\bar{1}0)$ plane perpendicular to the interface, by setting two of the four nearest-neighbor hoppings to be identical: $t_{\boldsymbol{\tau}_{2}}=t_{\boldsymbol{\tau}_{4}}=t_{0}$, $t_{\boldsymbol{\tau}_{1}}=t_{1}$, $t_{\boldsymbol{\tau}_{3}}=t_{2}$ where $t_{ij}=t_{\boldsymbol{\tau}_{\alpha}}$ with $\boldsymbol{\tau}_{\alpha}$ being the vector from $j$-th site to $i$-th site.
Topological classification based on mirror symmetry is possible by the mirror Chern number \cite{Teo08}, defined as
$n_{\cal M}=(n_{+i}-n_{-i})/2$.
Here $n_{\pm i}$ are the Chern numbers \cite{Thouless1982} within the subspace labeled with the mirror eigenvalues $\pm i$ in the mirror plane, 
and this Chern number corresponds to the number of chiral 
gapless modes within each subspace.
Because of the TRS, the mirror Chern number 
can also be written as $n_{\cal M}=n_{+i}=-n_{-i}$
 
If the wavevector is restricted to be within the mirror plane, $k_{x}=k_{y}=q$,
the mirror operator $\mathcal{M}$ can be diagonalized
by a unitary operator $V=\frac{1}{\sqrt{2}}\begin{pmatrix}
-\mathrm{e}^{i\frac{\pi}{4}}&\mathrm{e}^{i\frac{\pi}{4}}\\
1&1
\end{pmatrix}$ acting on a spin space. 
With the same matrix $V$, the model Hamiltonian (\ref{eq:FKMH}) can be transformed into a block-diagonal form
 $H_{\mathcal{M}}=\mathrm{diag}(H_{+i},H_{-i})$, 
where the Hamiltonian 
in the subspace $\mathcal{M}=\pm i$ is given by 
\begin{eqnarray}
H_{\eta i }&=&
\begin{pmatrix}
-2\sqrt{2}\eta\lambda f_{1}&f_{0}\\
f_{0}^{*}&2\sqrt{2}\eta\lambda f_{1}
\end{pmatrix}, \ \eta=\pm
\label{eq:etaHamiltonian},
\end{eqnarray}
where $
f_{0}(\mathbf{k})=t_{2}+2t_{0}\exp\left( -i\frac{q-k_z}{2}\right)+t_{1}\mathrm{e}^{-i q}
$ and $f_1(\mathbf{k}) 
=\sin \left( \frac{q+k_z}{2}\right)-\sin q+\sin \left( \frac{q-k_z}{2}\right)$.
Because $H_{ +i}(\lambda)=H_{-i}(-\lambda)$, the sign of $\lambda$ corresponds to the mirror Chern number $n_{\cal M}$.
Indeed, the mirror Chern number is calculated to be nontrivial, $n_{\mathcal{M}}=-{\rm sgn}(\lambda)$, when the system is in the TI phase \cite{Fu07}; namely, surface states of the 
model in the TI phase are protected not only by the 
TRS but also by the mirror symmetry\cite{Takahashi11,Rauch14}.
%

\begin{figure}[h]
 \begin{center}
 \includegraphics[width=80mm]{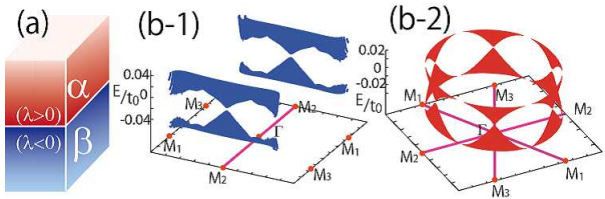}
 \caption{(a) Schematic illustration 
of the interface between two regions ($\alpha $ and $\beta $) of the FKM model .
(b-1,b-2) Interface energy bands near the Fermi energy ($E=0$) in the interface BZ for $\lambda_{\alpha }=0.15t_0$, and $\lambda_{\beta }=- 0.1t_0$. 
 (b-1) For $t_1=t_0$ and $t_2=1.4t_0$, the system does not have the rotational symmetry. There are two Dirac cones. (b-2) For $t_2=t_0$, $t_1=1.4t_0$, there are six Dirac cones due to C$_{3}$ symmetry.
}
\label{fig:Einterface}
\end{center}
\end{figure}

By making an interface between two regions $\alpha$, $\beta$ with the 
SOC parameters $\lambda_{\alpha }$ and $\lambda_{\beta }$ of opposite signs (Fig.~\ref{fig:Einterface}(a)),
gapless interface states are expected from the difference of the mirror Chern numbers. 
The resulting interface states are shown in Fig.~\ref{fig:Einterface}(b-1,b-2) for $(\lambda_{\alpha }, \lambda_{\beta })=(0.15,-0.1)t_0$, (b-1) $(t_1,t_2)=(1,1.4)t_0$, and (b-2) $(t_1,t_2)=(1.4,1)t_0$.
In both of the results,
two of the Dirac points are located on the mirror plane ($M_2$-$\Gamma$-$M_2$ line) in the 2D Brillouin Zone (BZ). This agrees with an argument of the mirror Chern number; the difference of $n_{\mathcal{M}}$ between the two regions is
$1-(-1)=+2$, implying that there are two right-going modes with ${\mathcal{M}}=+i$ and two 
left-going modes with ${\mathcal{M}}=-i$.
In (b-2), due to the C$_3$ symmetry around
[111], there are six Dirac cones, which
is in accordance with the low-energy model with $C_{3}$ warping terms \cite{Takahashi11}.


%

\begin{figure}[htbp]
 \begin{center}
 \includegraphics[width=85mm]{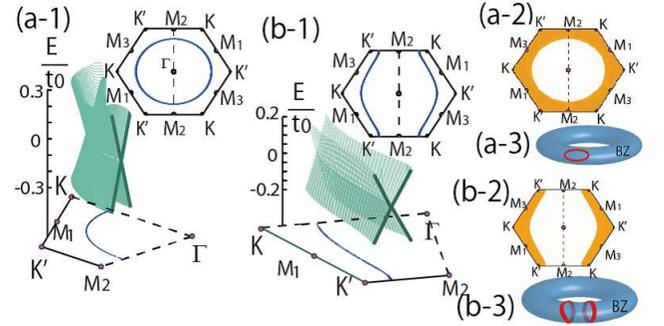}
 \caption{ (a-1, b-1) Dispersion and the Fermi surface for the interface states with IPHS.
Thickness of each region $\alpha$, $\beta$ is 24 unit cells along the [111] axis.
The parameters are
 (a) $(t_1, t_2) = (1.4, 1)t_{0}$, (b)$(t_1, t_2)=(1, 0.6)t_{0}$, and $\lambda_\alpha=-\lambda_\beta=0.125t_{0}$ for (a) and (b).
In the BZ in each panel, the Fermi surfaces at $E = 0$ are shown as blue curves.
 (a-2, b-2) Sign of the Pfaffian $\mathrm{sgn(pf}(\Tilde{U}H))$ in the interface BZ.
The shaded regions and white regions represent the negative and positive signs, 
respectively. 
Schematic illustrations of the Fermi loops of a closed orbit (a-3) and and 
open orbits (b-3) 
are shown on the torus of the interface BZ. 
}
\label{fig:PHSf}
\end{center}
\end{figure}

Next, we show that when $\lambda_\alpha=-\lambda_\beta$, a novel dispersion of interface states appears near the Fermi energy ($E=0$), as shown in Fig.~\ref{fig:PHSf}, 
which is in contrast with Fig.~\ref{fig:Einterface}.
The parameters of the SOC are $\lambda_\alpha=-\lambda_\beta = 0.125t_{0}$ for all the results, and $(t_1,t_2)$ is given as \ref{fig:PHSf}(a-1) $(1.4, 1)t_{0}$, 
and \ref{fig:PHSf}(b-1)$(1, 0.6)t_{0}$.
Figure~\ref{fig:PHSf}(a-1) belongs to the TI phase, and \ref{fig:PHSf}(b-1) is in the non-TI phase.
 In all the cases, the gap closes at $E=0$, with its degeneracy forming loops in the interface BZ. 
The dispersion is linear in the direction perpendicular to the loop. 
Remarkably, this loop is not on a high-symmetry line.
This kind of degeneracy along a line is unexpected in general lattice models, because there is no symmetry which protects 
degeneracy between the valence and conduction bands along a loop.
Therefore, 
new theoretical description is required.

For the sake of explaining this peculiar dispersion, we consider a transformation of the FKM Hamiltonian as
\begin{eqnarray}
UH_{\lambda,\mathbf{k}}U^{-1}=-H^{t}_{-\lambda,-\mathbf{k}},\label{eq:lamrelation}
\label{eq:UHU}
\end{eqnarray}
where $U=i\tau_{z}s_{y}$. 
The Hamiltonian also has an inversion symmetry: $PH_{\lambda,\mathbf{k}}
P^{-1}=H_{\lambda,-\mathbf{k}},\ P=\tau_x$. An operator $UK=i\tau_z s_yK$ corresponds to 
the following sublattice-dependent PH transformation:
$\Tilde{c}^{\dagger}_{\alpha i \pm}=\pm c_{\alpha i\mp}$,
$\Tilde{c}^{\dagger}_{\beta i\pm}=\mp c_{\beta i\mp}$,
which leaves the spin operator invariant: $c_{i}^{\dagger}\mathbf{s} c_{i}=\Tilde{c}_{i}^{\dagger}\mathbf{s}\Tilde{c}_{i}$.
The respective regions $\alpha$ and $\beta$ are not invariant under the PH transformation, but they are transformed to each other 
by this PH transformation because $\lambda_\alpha=-\lambda_\beta$. Therefore, by making the $\alpha$-$\beta$ interface,
the PHS is restored as shown in the following. 

The existence of the anomalous interface states forming a Fermi loop is explained as 
follows.
First we note that for a Hamiltonian $H^{\rm{2D}}_{\lambda ,\mathbf{k}_{\|} }$ describing a
2D slab with finite thickness, 
we have 
$UH^{\rm{2D}}_{\lambda ,\mathbf{k}_{\|} }U^{-1}= -
(H^{\rm{2D}}_{-\lambda ,-\mathbf{k}_{\|} })^{t}$
from Eq.~(\ref{eq:UHU}), where $\mathbf{k}_{\|}$ is a 2D wavevector parallel to the interface.
Then the junction of two regions $\alpha, \beta$ (Fig.~\ref{fig:Einterface}(a)),
with their SOI parameters given by $\lambda_\alpha=-\lambda_\beta=\lambda$ is described by the Hamiltonian
 \begin{eqnarray}
 H_{\mathbf{k}_{\|}}=
 \begin{pmatrix}
 H^{\rm{2D}}_{\lambda ,\mathbf{k}_{\|} }&V_{\mathbf{k}_{\|}}\\
 V^{\dagger}_{\mathbf{k}_{\|}}&H^{\rm{2D}}_{-\lambda,\mathbf{k}_{\|}}
 \end{pmatrix},
\end{eqnarray}
where $V$ is the hybridization between the regions $\alpha$ and $\beta$.
We have the following relation,
 \begin{eqnarray}
 U
 H_{\mathbf{k}_{\|}}
 U^{-1}
&=& \begin{pmatrix}
-(H^{\rm{2D}}_{-\lambda,-\mathbf{k}_{\|} })^{t}&U V_{\mathbf{k}_{\|}}U^{-1}\\
 UV_{\mathbf{k}_{\|}}^{\dagger} U^{-1}& -(H^{\rm{2D}}_{\lambda ,-\mathbf{k}_{\|} })^{t}
 \end{pmatrix}.
\end{eqnarray}
Therefore, the SOC parameter $\lambda$ changes sign, and 
the regions $\alpha$ and $\beta$ are exchanged. 
When the two regions $\alpha$, $\beta $ have the same thickness, 
we can exchange the regions again by the space inversion. The space inversion is
given by $PP'\Sigma_x$, where
$P'$ is an operator for inversion of the stacking direction of layers within each region
and $\Sigma_{x }$ is the Pauli matrix acting to the region degrees of freedom, $\alpha$ and $\beta$.
When each region $\alpha$ and $\beta$ have $N$ AB pairs along the stacking direction, $P'$ is given as $P'_{ij}=\delta_{i,N+1-j}\tau_{0}s_{0}$, where 
$i$ and $j$ denote the indices for the sublattice pair from the top, and $\tau_{0}$ and $s_{0}$ are the identity operators for the sublattice and spin spaces.
Therefore when the hybridization $V$ satisfies
 \begin{eqnarray}
P P'UV_{\mathbf{k}_{\|}} (PP'U)^{-1}=-V^{t}_{\mathbf{k}_{\|}}, 
\label{eq:Vcondition}
\end{eqnarray}
two regions $\alpha$ and $\beta$ are exchanged by the space inversion,
and the PHS is restored:
 \begin{eqnarray}
\tilde{U}H_{\mathbf{k}_{\|}} \tilde{U}^{-1}=-H_{\mathbf{k}_{\|}}^{t} \label{eq:PHUK}, 
\end{eqnarray}
where 
$\Tilde{U}=Q\Sigma_{x}$ and $Q=PP'U$. 
Thus, although each Hamiltonian is not invariant under a combined transformation of $U$ and the inversion, by making 
a junction consisting of two regions $\alpha $, $\beta $, the system restores a symmetry under a PH transformation $U$ and the spatial inversion, 
which we call {\it the interfacial particle-hole symmetry (IPHS)}.
Thus the energy eigenvalues of the whole system are symmetric with respect to $E=0$.

This new symmetry gives rise to emergence of the Fermi loop in the interface states, as we show in the following.
From Eq.~(\ref{eq:PHUK}), $\Tilde{U} H_{\mathbf{k}_{\|}}$ is a skew symmetric matrix,
 \begin{eqnarray}
[\Tilde{U} H_{\mathbf{k}_{\|}}]^{t}&=& H_{\mathbf{k}_{\|}}^{t}\Tilde{U}^{t}= -\Tilde{U}H_{\mathbf{k}_{\|}} \Tilde{U}^{-1} \Tilde{U}^{t}=-\Tilde{U}H_{\mathbf{k}_{\|}},
\end{eqnarray}
and furthermore its Pfaffian is real: 
\begin{eqnarray}
\mathrm{pf}[\Tilde{U} H_{\mathbf{k}_{\|}}]&=&\mathrm{pf}[-H^{t}_{\mathbf{k}_{\|}}\Tilde{U}^{t} ]=
\mathrm{det}(\tilde{U})\mathrm{pf}[\Tilde{U}^{-1}H^{t}_{\mathbf{k}_{\|}} ]
\nonumber\\
&=&\mathrm{pf}[\Tilde{U}H_{\mathbf{k}_{\|}} ]^{*}.
\end{eqnarray}
To derive these relations, we used the fact that the matrix $\tilde{U}$ is unitary and symmetric, and the fact that the size of the matrix is an integer multiple of four.
The reality condition of $\mathrm{pf}[\Tilde{U} H_{\mathbf{k}_{\|}}]$ gives a strong constraint for the Fermi surface.
Suppose the sign of $\mathrm{pf}(\Tilde{U} H_{\mathbf{k}_{\|}})$ changes at some $\mathbf{k}_{\|}$; 
it then means $\mathrm{pf}(\Tilde{U} H_{\mathbf{k}_{\|}})=0$, i.e. $\mathrm{det}\Tilde{U}
\det H_{\mathbf{k}_{\|}}=0$ and $\mathrm{det}H_{\mathbf{k}_{\|}}=0$. Therefore, the 
Hamiltonian $H_{\mathbf{k}_{\|}}$ has zero eigenvalues and the gap closes at this wavevector $\mathbf{k}_{\|}$, because of the PH symmetry.
Because the wavevector $\mathbf{k}_{\|}$ where the sign of $\mathrm{pf}(\Tilde{U} H_{\mathbf{k}_{\|}})$ changes forms a loop in the 2D BZ, this loop 
is nothing but the Fermi loop where the gap is closed. 
We note that we have set the thickness of the two regions $\alpha$ and $\beta$ to
be the same. This can be relaxed
as long as the 
slab thickness is much larger than the decay length of the interface states.

%
To confirm that the Fermi loop shown in Fig.~\ref{fig:PHSf} is due to this scenario, 
we calculate the Pfaffian $\mathrm{pf}(\Tilde{U} H)$ in this model by using PFAPACK \cite{Wimmer12}.
The results are shown in Figs.~\ref{fig:PHSf}(a-2) and \ref{fig:PHSf}(b-2) by using the same parameters for \ref{fig:PHSf}(a-1) and \ref{fig:PHSf}(b-1).
Here the shaded and white regions represent
$\mathrm{sgn(pf}(\Tilde{U}H_{\mathbf{k}_{\|}}))=-1$, and $\mathrm{sgn(pf}(\Tilde{U}H_{\mathbf{k}_{\|}}))=1$,
respectively. Hence the wavevectors $\mathbf{k}$ for the sign change of
 $\mathrm{pf}(\Tilde{U} H_{\mathbf{k}_{\|}})$ agree with the
interface Fermi loops shown in Fig.~\ref{fig:PHSf}.
In addition, the Fermi loops in (a-2) form closed orbits, while those in (b-2) do open orbits.
These behaviors are similar to the form of the Fermi surface without the SOC \cite{Takahashi13}.
Due to the reality of $\mathrm{pf}(\Tilde{U} H)$, the Fermi loops are classified into two types: closed orbits and open orbits (Fig.~\ref{fig:PHSf}(a-3) and (b-3)). 
Because of the nontrivial winding number for the open orbits, 
it is not possible for the open-orbit Fermi loops to disappear without transition to closed loops.
%


To realize gapless topological 
interface states due to the mirror Chern number experimentally, two TIs with opposite chiralities are necessary and mirror symmetry normal to the interface is required in common.
In a recent experiment, 
 a natural superlattice, Bi$_4$Se$_{2.6}$S$_{0.4}$ is reported to have
a surface Dirac cone with the chirality opposite from the conventional one
 \cite{Valla12}, although it is not a TI but a semimetal.
Furthermore, it might be interesting to search for interface Fermi loops between two insulators from this scenario.

Our theory on topological Fermi loops can be generalized. We construct a system
 with two regions $\alpha, \beta$ (Fig.~\ref{fig:Einterface}(a)), 
and they are assumed to be related to each other by an operator $U$: 
$H_{\alpha(\beta),\mathbf{k}_{\|}} = \varepsilon_{C}UH^{t}_{\beta(\alpha),-\mathbf{k}_{\|}}U^{-1}$, where 
$\varepsilon_{C}=\pm1$. We can classify the operator $U$ as PHS-like for $\varepsilon_{C}=-1$ or TRS-like for $\varepsilon_{C}=1$.
As is similar to the FKM model, if the hybridization $V$ between the two regions satisfies the condition,
$\tilde{P}UV_{\mathbf{k}_{\|}} (\tilde{P}U)^{-1}=\varepsilon_{C}V^{t}_{\mathbf{k}_{\|}}
$, where $\tilde{P}$ stands for the space inversion 
for a finite-thickness slab, the Hamiltonian satisfies a relation
 \begin{eqnarray}
\tilde{U}H_{\mathbf{k}_{\|}} \tilde{U}^{-1}=\varepsilon_{C}H_{\mathbf{k}_{\|}}^{t},\ 
\tilde{U}\equiv \tilde{P}U\Sigma_x.
\end{eqnarray}
Thus, by making 
a junction between the regions $\alpha $ and $\beta$, the system restores a symmetry under a combined transformation of $U$ and inversion.
 We call this symmetry IPHS for $\varepsilon_{C}=-1$ as mentioned previously, and {\it interfacial time-reversal symmetry (ITRS)} for $\varepsilon_{C}=1$.
These new symmetries are classified by $\tilde{U}=\eta_{U}\tilde{U}^{t},\ \eta_U=\pm1$.
Because 
$[\Tilde{U} H_{\mathbf{k}_{\|}}]^{t}= H_{\mathbf{k}}^{t}\Tilde{U}^{t}= \varepsilon_{C}\eta_{U}\Tilde{U}H_{\mathbf{k}_{\|}}$, 
$\Tilde{U} H_{\mathbf{k}}$ is a skew symmetric matrix only for $\varepsilon_{C}\eta_{U}=-1$, i.e. $(\varepsilon_{C}, \eta_{U})=(\mp 1,\pm 1)$. Then, the Pfaffian $\mathrm{pf}(\Tilde{U} H_{\mathbf{k}_{\|}})$
is defined and is real as before. 

The reality of the Pfaffian $\mathrm{pf}(\Tilde{U} H_{\mathbf{k}_{\|}})$
is reflected in the dispersions of interface states in a different way between 
the IPHS and the ITRS. 
For junction systems with the IPHS ($(\varepsilon_{C}, \eta_{U})=(-1,1)$) between two insulating regions, the gap between the valence and conduction bands closes along a Fermi loop, as we saw in the FKM model. 
In this case, the two regions are related by the TRS, and have different bulk band structure. 
It is usually unlikely for the gap to close at an interface between two insulators, and it is even more unlikely to close the gap along a loop. Nevertheless, for the present cases with the IPHS, if the gap of the interface states
is closed somewhere,
the gap should close not at an isolated point but along a loop due to the reality of the Pfaffian. 
In the example of the FKM model, there is a reason for the gap to close; namely the difference of the mirror Chern number between the two regions results in the presence of 
gap-closing points 
along the mirror plane. Together with the reality of the Pfaffian, the gap closes along a loop. 

In fact the appearance of Fermi loops in the IPHS class is not limited to cases with topological interfacial states (from the mirror Chern number), but can be found in general junction systems with IPHS, as long as 
the hybridization at the interface is stronger than the gap. 
To show this, we consider a junction system with weak hybridization $V_{\mathbf{k}_{\|}}$ between 
two regions $\alpha$ and $\beta$.
We diagonalize the Hamiltonian for the 
$\alpha$-region, $H_{\alpha ,\mathbf{k} _{\|}}$, by a unitary matrix $W_{\mathbf{k} _{\|}}$ as $W_{\mathbf{k} _{\|}}^{\dagger}H_{\alpha ,\mathbf{k} _{\|}}W_{\mathbf{k} _{\|}}=\mathrm{diag}(E_{1\mathbf{k}_{\|}},E_{2\mathbf{k}_{\|}},...,E_{\bar{N}\mathbf{k}_{\|}})$ with $E_{i\mathbf{k}_{\|}}<E_{i+1\mathbf{k}_{\|}}$ for $i=1,2,...,\bar{N}-1$, and $\bar{N}$ is 
the size of the matrix.
The two regions are band insulators, and the Fermi energy $E_{F}(=0)$ is assumed to be between $E_{m\mathbf{k}_{\|}}$ and $E_{m+1\mathbf{k}_{\|}}$. By the interface transformation, the Hamiltonian for the $\beta$-region 
$H_{\beta ,\mathbf{k}_{\|} }$ is diagonalized by
 $W'_{\mathbf{k} _{\|}}\equiv U^{*}W_{-\mathbf{k}_{\|}}^{*}$ as $W_{\mathbf{k} _{\|}}^{\prime\dagger}H_{\beta,\mathbf{k} _{\|}}W'_{\mathbf{k} _{\|}}=\mathrm{diag}
(-E_{1,-\mathbf{k}_{\|}},-E_{2,-\mathbf{k}_{\|}},...,-E_{\bar{N},-\mathbf{k}_{\|}})$.
By assuming 
$|E_{m\mathbf{k}_{\|}}|,|E_{m+1\mathbf{k}_{\|}}| \ll |E_{i\mathbf{k}_{\|}}|$ for $i\not=m,m+1$, 
leading-order terms of the Pfaffian expanded in terms of $V'$ are given as
 \begin{eqnarray}
\mathrm{pf}(\Tilde{U}H_{\mathbf{k}_{\|}})\sim Z_{\bar{N}}\left[1+\frac{|V'_{m,m+1,-\mathbf{k}_{\|}}|^2 }{E_{m,-\mathbf{k}_{\|}}E_{m+1,-\mathbf{k}_{\|}}}\right],
\end{eqnarray}
where $Z_{\bar{N}} = \prod_{i=1}^{\bar{N}}E_{i,-\mathbf{k}_{\|}}$.
Since each region is assumed to be gapped for the entire BZ, the sign of $Z_{\bar{N}} $ is fixed. Thus, the condition for the Fermi loop is given as
 \begin{eqnarray}
E_{m\mathbf{k}_{\|}}E_{m+1\mathbf{k}_{\|}} =-|V'_{m,m+1,\mathbf{k}_{\|}}|^2 \label{eq:FLcond}.
\end{eqnarray}
It means that the Fermi loop appears when 
the hybridization becomes comparable to the gap for the IPHS.
For example, when the chemical potential to be the middle of the gap ($-E_{m+1,\mathbf{k}_{\|}}=
E_{m,\mathbf{k}_{\|}}$), the Fermi loop appears at
$E_{m+1,\mathbf{k}_{\|}}
= |V'_{m,m+1,\mathbf{k}_{\|}}|$. 

On the other hand, for the ITRS class $(\varepsilon_{C}, \eta_{U})=(1,-1)$, 
it results in double degeneracy for every eigenstate,
because the eigenequation becomes a perfect square: $\mathrm{Det}(H-E)=\mathrm{Det}(\tilde{U}(H-E))=(\mathrm{pf}(\tilde{U}(H-E)))^2$. Here we used the fact that $\tilde{U}$ is 
skew symmetric for the ITRS class. Therefore, although within each region the TRS symmetry is broken, the ITRS gives rise to a ``Kramers-like'' degeneracy for every state. 


In general, it might be difficult to realize interfaces with the IPHS.
On the other hand, it is easier for superconductors, because their energy bands have the PHS by themselves.
Here we show another example of the Fermi loop at the $\pi$-junction interface between two 2D Rashba 
systems.
We consider 2D bilayer systems with the Rashba SOC on the square lattice \cite{Bernardes07,Akabori12}, 
and assume that s-wave superconductivity is induced by a proximity effect.
In the case with the monolayer Rashba system, the Bogoliubov-de Gennes Hamiltonian is written as
 \begin{eqnarray}
 H_{\Delta}(\mathbf{k})&=&
 \begin{pmatrix}
 h_{0}(\mathbf{k})&is_{y}\Delta\\
 -is_{y}\Delta& -h_{0}^{t}(-\mathbf{k})
 \end{pmatrix},\\
 h_{0}(\mathbf{k})&=&\xi(\mathbf{k})+\alpha(s_{x}\sin k_{y}-s_{y}\sin k_{x}),
\end{eqnarray}
where $\Delta$ is the superconducting order parameter and is assumed to be a real constant, $\xi(\mathbf{k})$ ($=\xi(-\mathbf{k})$) is the kinetic energy from the chemical potential $\mu$, and $\alpha$ is the strength of the SOC.
The eigenvalues are $E_{\Delta}=\pm\sqrt{ \left(\xi(\mathbf{k})\pm\alpha\sqrt{\sin^2 k_{x}+\sin^2 k_{y}}\right)^2+\Delta^2 }$, and the system is gapped for the entire BZ.
Because the Hamiltonian satisfies the equations
$ \tau_{y}H_{\Delta} (\mathbf{k})\tau_{y}=-H_{-\Delta}^{t} (-\mathbf{k})$, and 
$ \tau_{z}s_{z}H_{\Delta} (\mathbf{k})\tau_{z}s_{z}=H_{\Delta} (-\mathbf{k})$, we can see that 
the Hamiltonians $H_{\pm\Delta}(\mathbf{k})$ are related by the PHS with $(\varepsilon_{C},\eta_{U})=(-1,1)$.
Therefore, 
a $\pi$-junction interface between two 2D Rashba systems with $+\Delta$ and $-\Delta$ has the IPHS, and its Hamiltonian is given by
 \begin{eqnarray}
 H(\mathbf{k})&=&
 \begin{pmatrix}
 H_{\Delta} (\mathbf{k})&t\tau_{z}\\
 t\tau_{z}& H_{-\Delta} (\mathbf{k})
 \end{pmatrix},
\end{eqnarray}
where $t$ is the hybridization across the junction, and $\tau_{z}$ is the Pauli matrix for Nambu space.
At the band edge, we approximate the kinetic energy as $\xi(\mathbf{k})=\frac{k^2}{2m_{e}}-\mu$, where $m_{e}$ is the electron mass, 
and $\sin k_i\sim k_i$. We, then, obtain the energy eigenvalues
$E_{\Delta}=\pm t\pm\sqrt{ [\frac{k^2}{2m_{e}}-\mu\pm\alpha k]^2+\Delta^2 }$.
Therefore, 
the Fermi loops appear when the hybridization is larger than the gap ($|t|>|\Delta|$).
It can be satisfied even if the junction is weak, 
because the hybridization $t$ can be maximally $\sim 10^{-1}$ eV, and the gap $\Delta$ is typically 
 $\Delta \sim 10^{-4}$eV.

In conclusion,
we found new symmetries called interfacial symmetries, IPHS and ITRS, in junction systems, and these symmetries appear as novel dispersions of
interface gapless states. 
Here the respective regions on either side of the junction are not invariant under a particular transformation, but the 
whole junction system is invariant. 
The interface Fermi loop is guaranteed by the reality of the Pfaffian of a skew-symmetric matrix: $\mathrm{pf}(\Tilde{U} H)$ where $\Tilde{U}$ consists of the space inversion and the PH transformation. We found that the Fermi loop is a boundary between the positive and negative regions of 
$\mathrm{pf}(\Tilde{U} H)$. 
This presence of the Fermi loop is shown for general systems with the IPHS. On the other hand, the ITRS leads to double degeneracy for every state in junction systems. 

This work is partially supported by the Global Center of Excellence Program by MEXT, Japan through the ``Nanoscience and Quantum Physics" Project of the Tokyo Institute of Technology, Grant-in-Aid from MEXT, Japan (No. 21000004 and 26287062), JSPS Re- search Fellowships for Young Scientists, Grant-in-Aid for JSPS Fellows No. 25-9798, and Tokodai Institute for Element Strategy (TIES) funded by MEXT Elements Strategy Initiative to Form Core Research Center.


\begin{thebibliography}{99}


\bibitem{Kane05a}C. L. Kane and E. J. Mele, Phys. Rev. Lett. {\bf 95}, 226801 (2005).


\bibitem{Moore07} J. E. Moore and L. Balents, Phys. Rev. B \textbf{75}, 121306(R) (2007).
 
 \bibitem{Fu07}L.~Fu, C.~L.~Kane and E. J. Mele, Phys. Rev. Lett. \textbf{98}, 106803 (2007).

 \bibitem{Hasan10}
M. Z. Hasan, C. L. Kane, Rev. Mod. Phys. \textbf{82}, 3045 (2010).
 
 
 
\bibitem{Hsieh09a}D. Hsieh {\it et al.}, Science \textbf{323}, 919 (2009).
\bibitem{Hsieh09b}D. Hsieh {\it et al.}, Nature \textbf{460}, 1101 (2009).



\bibitem{Xia09}Y. Xia {\it et al.}, Nature Phys. \textbf{5}, 398 (2009).
\bibitem{Chen09} Y. L. Chen {\it et al.}, Science \textbf{325}, 178 (2009).





\bibitem{Teo08} J. C. Y. Teo, L. Fu, C. L. Kane, Phys. Rev. B \textbf{78}, 045426 (2008).
\bibitem{THsieh12}
T. H. Hsieh {\it et al.}, Nat. Commun. {\bf 3}, 982 (2012).
\bibitem{Takahashi11}R. Takahashi, S. Murakami, Phys. Rev. Lett., {\bf 107}, 166805 (2011).



\bibitem{Thouless1982} D. J. Thouless, M. Kohmoto, M. P. Nightingale, and M. den Nijs, Phys. Rev. Lett. \textbf{49}, 405 (1982).
\bibitem{Rauch14}
T. Rauch, M. Flieger, J. Henk, I. Mertig, and A. Ernst, Phys Rev. Lett. \textbf{112}, 016802 (2014).



\bibitem{Wimmer12}
M. Wimmer, ACM Trans. Math. Software \textbf{38}, 30 (2012).

\bibitem{Takahashi13}
R. Takahashi, S. Murakami, Phys. Rev. B \textbf{88}, 235303 (2013).
 \bibitem{Valla12}
T. Valla {\it et al.}, Phys. Rev. B 86, 241101(R) (2012).

\bibitem{Bernardes07}
E. Bernardes, J. Schliemann, M. Lee, J. C. Egues, and D. Loss,
Phys. Rev. Lett. \textbf{ 99}, 076603 (2007).
\bibitem{Akabori12}
M. Akabori {\it et al.},J. Appl. Phys.\textbf{ 112}, 113711 (2012).





\end{thebibliography}

\end{document}